# Heuristics facilitates the evolution of transitive inference and social hierarchy in a large group

Kazuto Doi and Mayuko Nakamaru

Department of Technology and Innovation Management, Tokyo Institute of Technology
CIC-901S, 3-3-6 Shibaura, Minato-Ku, Tokyo, 108-0023, Japan

E-mail:
    K. Doi           83nsoundbeach@gmail.com
    M. Nakamaru    nakamaru.m.aa@m.titech.ac.jp

**Abstract:**

Transitive inference (TI) refers to social cognition that facilitates the discernment of unknown relationships between individuals using known relationships. It is extensively reported that TI evolves in animals living in a large group because TI could assess relative rank without deducing all dyadic relationships, which averts costly fights. The relationships in a large group become so complex that social cognition may not be developed adequately to handle such complexity. If all group members apply TI to all possible members in the group, TI is supposed to require extremely highly developed cognitive abilities especially in a large group. Instead of developing cognitive abilities significantly, animals may apply simplified TI we call reference TI in this study as heuristic approaches. The reference TI allows members to recognize and remember social interactions only among a set of reference members rather than all potential members. Our study assumes that information processes in the reference TI comprises 1) the number of reference members based on which individuals infer transitively, 2) the number of reference members shared by the same strategists, and 3) memory capacity. We examined how information processes evolve in a large group using evolutionary simulations in the hawk–dove game. Information processes with almost any numbers of reference members could evolve in a large group as long as the numbers of shared reference member are high because information from the others' experiences is shared. TI dominates immediate inference, which assesses relative rank on direct interactions, because TI could establish social hierarchy more rapidly applying information from others' experiences.

# 1. Introduction
## 1.1. Background

How to increase chances of winning competitions for limited resources is critical for animals living in groups (Austad, 1983; Enquist and Leimar, 1983; Milinski and Parker, 1991). The asymmetric hawk–dove framework has often been employed in the analysis of the evolution of fighting behavior in animals (Parker, 1974; Maynard Smith, 1974; Maynard Smith and Parker, 1976). In the hawk–dove game, players select a tactics between hawk (escalation) and dove (retreat) based on their inference strategies. In hawk vs. hawk, players with higher resource-holding potential (RHP) have a higher chance of winning a contest. If both select hawk, a winner gains a reward, and the loser incurs a loss. If both select dove, they gain half of the reward equally. If one player chooses hawk and the other chooses dove, the hawk wins the entire reward, and the dove receives nothing. Therefore, it is critical to assess the RHP of an opponent using historical performance in previous contests within a group (Enquist and Leimar, 1983). Previous theoretical and empirical studies have revealed that the assessment of RHP is applied in two different ways including the abilities to accurately assess RHP and to promptly form the social hierarchy (Arnott and Elwood, 2009; Hsu et al., 2006; Milinski and Parker, 1991; Parker, 1974; Reichert and Quinn, 2017). An accurate assessment increases the chances of winning, whereas the prompt formation of the social hierarchy averts costly fights (Maynard Smith, 1974; Mesterton-gibbons and Dugatkin, 1995; Smith and Price, 1973). Nakamaru and Sasaki (2003) theoretically demonstrated that the ability to accurately assess RHP is favored when the cost of losing is relatively low because the hawk vs. hawk combination that occurs more often with lower costs provides useful information on relative RHP because hawk vs. hawk leads to actual fights. In contrast, the ability to form the social hierarchy promptly would be favored more when the cost is relatively high. The former ability to make an accurate assessment is demonstrated in a strategy referred to as immediate inference (II) strategy in Doi and Nakamaru (2018) and Nakamaru and Sasaki (2003) where a player who estimates the strength of an opponent based on the history of direct fights. The latter ability to quickly build the social hierarchy is associated with transitive inference (TI) strategy, which estimates the strength of an unknown by using known relationships (Doi and Nakamaru, 2018; Nakamaru and Sasaki, 2003). Transitive inference is useful when A knows that A is stronger than B and B is stronger than C, but has no idea about the relationship between A and C. If A has the ability for transitive inference, A could infer A > C, using A > B and B > C even when A has never interacted with C. Immediate inference demonstrates the ability of accurate assessment while transitive inference proves the ability of the prompt formation of the social hierarchy (Doi and Nakamaru, 2018; Nakamaru and Sasaki, 2003). Both types of inferences have been reported extensively in the animal kingdom (Allen, 2013; Grosenick et al., 2007; Paz-Y-Miño et al., 2004; Vasconcelos, 2008; White and Gowan, 2013).

Both immediate and transitive inferences require social cognition, which refers to information learned about the characteristics of other individuals in the course of social interactions or based on observations (Sheehan and Bergman, 2016). However, social cognition required by different types of inferences is considerably different. For example, immediate inference requires individuals to recognize only individuals that they have interacted with while transitive inference requires individuals to recognize a much broader range of individuals regardless of whether they have interacted or not (Bshary and Brown, 2014; Lilly et al., 2019; Seyfarth and Cheney, 2015). Social cognition has been investigated extensively in a wide range of animals, including both vertebrates and invertebrates (Emery et al., 2007; Gheusi et al., 1994). In the present study, we consider social cognition as a set of processes to recognize others broadly regardless of direct or indirect interactions and recall information about others. Social cognition in transitive inference includes the ability to observe and remember social interactions among others as well as own interactions; in contrast, social cognition in immediate inference is limited to the ability to observe and remember own social interactions and does not involve the observation and memory of the interactions of others.

The social complexity hypothesis suggests that living in large social groups facilitates the evolution of cognitive abilities (Balda and Kamil, 1989; Fernald, 2014, 2017; Jolly, 1966; MacLean et al., 2008; Waal and Tyack, 2003). According to this hypothesis, societies where the social hierarchy is critical could promote the evolution of social cognition. For example, the number of members in a group in the study on the social hierarchies in *Astatotilapia burtoni* was 20 (Fernald, 2014). In addition, Reichert and Quinn (2017) and Hobson (2020) highlighted the importance of cognitive mechanisms that drive contest behaviors. However, little is known about such cognitive mechanisms. Transitive inference is considered to evolve in animals living in large groups as a way of facilitating the understanding of the social hierarchy without increasing memory capacity when the number of dyadic relationships significantly increases with an increase in the group size (Mikolasch et al., 2013; Paz-Y-Miño et al., 2004). A recent report that transitive inference is observed even in insects such as wasps adds evidence that the miniature nervous system of insects does not limit sophisticated social behaviors (Tibbetts et al., 2019).

Doi and Nakamaru (2018) studied the two types of inference, immediate and transitive inferences, in the asymmetric hawk–dove game, in light of the impacts to inferences by limited memory capacity, by analyzing the evolutionary dynamics using computer simulations. They revealed that transitive inference evolves with relatively low memory capacity when the cost of losing in the hawk–dove game is relatively high. The reason is that transitive inference can form the social hierarchy promptly even with relatively low memory capacity. Lower memory capacity is even more effective because lower memory capacity enhances the consistency of the social hierarchy with ranking based on RHP by disregarding existing social hierarchy that is inconsistent with RHP and adjusting the hierarchy through actual fights resulting from hawk vs. hawk interactions. It is important to note that the social hierarchy built rapidly using transitive inference does not necessarily represent the actual RHP rank appropriately.

Theoretical findings above by Doi and Nakamaru (2018) support the social complexity hypothesis. However, transitive inference in their models assumed highly developed social cognition that allowed individuals in a group to recognize any other individuals that the individual had not interacted with and remembered all the outcomes of contests among the individuals. Hereafter, the group size is abbreviated as *N*. Under this assumption, the relationships among members become increasingly complex and information required for understanding the social hierarchy significantly increases, as the group size increases. If all group members apply transitive inference to all possible members in group, transitive inference is supposed to require extremely highly developed cognitive abilities especially in a large group where all members recognize all other members, observe and remember all social interactions among all members. Therefore, the assumption of extremely highly developed cognitive abilities in Doi and Nakamaru (2018) could be too impractical with an increase in the group size, so that we relax this assumption in our present study. Instead of developing cognitive abilities substantially, animals may apply a kind of simplified transitive inference we call reference transitive inference in this study as heuristic approaches. The reference transitive inference allows members to recognize and remember social interactions only among a set of reference member rather than all potential members. Animals may apply some shortcuts, or heuristic approaches such as reference transitive inference to handle such complex scenarios, instead of developing social cognition accordingly. Therefore, in the present study we consider situations where individuals in a group apply information based only on relationships with some members, which we name a set of reference members, as opposed to that based on all potential members.

## 1.2. *Heuristics and social complexity*

We consider social cognition as a set of processes to a) make an inference and b) to gather and store the information for inference. The first part is referred to as inference processes while the second part is referred to as information processes (e.g., Table 1) in this study. Inference processes consist of immediate inference and transitive inference, while information processes comprise three parts: 1) The number of reference members, which represents the number of members who the focal individual can recognize and focus on, 2) The number of reference members shared by individuals (Fig. 1) and 3) Memory capacity. Two former parts, 1) and 2), in information processes, correspond to heuristic mechanisms in transitive inference.

We refer to a set of individuals in a group as a set of reference members based on which individuals infer transitively. A set of reference members can be a group of arbitrary members. Through the set of reference members, transitive inference players can apply information from experiences by other reference members. On the other hand, immediate inference players can use information only from experiences by themselves. In addition, we assume the ability to share reference members with individuals following the same strategy (Fig. 1). Sharing reference members enables members opting for the same strategy to share information based on the experiences of shared reference members in a group. As a result, sharing reference members promotes the formation of the social hierarchy. In the present study, we redefine transitive inference as $TI_{x-y}$, where individuals can recognize and focus on an $x$ number of reference members ($x \leq N - 1$). Individuals following the same strategy share $y$ number of members out of $x$ number of reference members ($y \leq x \leq N - 1$). Figure 1 shows that how the number of reference members and the number of shared reference members in the transitive inference-process interact. Player A and C both employ a $TI_{3-2}$ strategy. Considering the number of reference members = 3, we assume that the reference members for A are D, E and F, and the reference members for C, are D, E and H. Players applying the $TI_{3-2}$ strategy are assumed to share two players D and E with other $TI_{3-2}$ players. D and E are shared reference members for all $TI_{3-2}$ players in the group. Shared reference members would help us to understand how sharing the same information in TI promotes the formation of the social hierarchy.

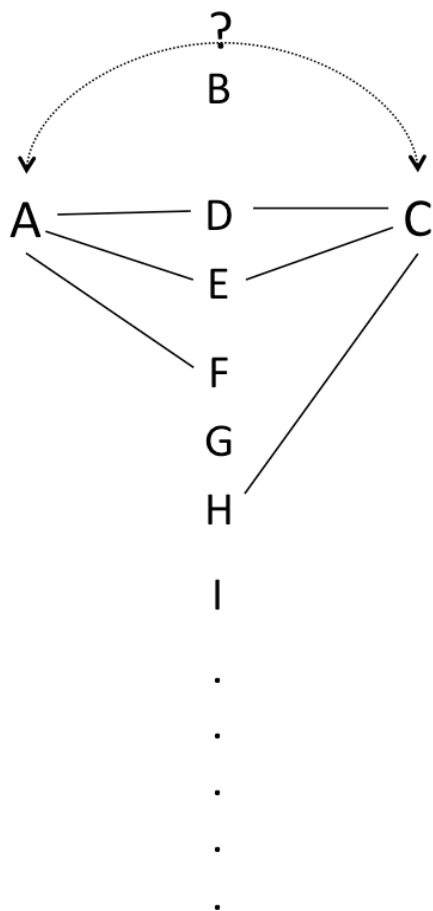

Fig. 1: Transitive inference process, the number of reference members and the number of shared reference members in $TI_{X-Y}$ in the case of two $TI_{3-2}$ players in a group

Players A and C follow the $TI_{3-2}$ strategy. Given the number of reference members = 3, solid lines show that reference members for A are D, E and F and reference members for C, are D, E and H. Players applying the $TI_{3-2}$ strategy are assumed to share two players with the other $TI_{3-2}$ players since the number of shared reference members is 2. Shared reference members for A and C are D and E in the present example. Reference members are randomly chosen. Dotted line shows that A and C attempt to make an assessment of the relative rank each other using transitive inference-process when there are no direct contests between A and C. Player B adopts the immediate inference strategy.

The heuristic approaches in transitive inference, or the ability to have a set of reference members and share reference members, could substantially reduce the number of pairs required for understanding the entire dominance hierarchy, compared to immediate inference, particularly when the group size is large. For example, immediate inference needs information about $N \times (N-1)/2$ pairs while $TI_{1-1}$ requires only $N-1$ relationships at minimum to understand the entire dominance hierarchy in a group. Even limited number of reference members and shared reference members could facilitate the establishment of the social hierarchy rapidly.

The social complexity hypothesis suggests that survival in a large group requires the ability to form the social hierarchy promptly (Bond et al., 2003, 2010; Seyfarth and Cheney, 2003, 2015). Social complexity is a common, but a little controversial, concept due to a lack of objectivity and a failure to link sociality to the application of cognition (Bergman and Beehner, 2015). A review study about goldfish and parrots by Croney and Newberry (2007) and a comparative study of six primate species by MacLean et al. (2013) suggest that the group size signficantly influences the development of social cognition. However, the use of the group size as an index of social complexity is sometimes criticized because it does not take into account the diverse interactions among different animals within groups (Bergman and Beehner, 2015). In the present study, we consider the size of a social group one of components influencing social complexity for the sake of simplicity.

## 2. Model
### 2.1. Hawk-dove game

We consider a group of $N$ players. We pick two players, A and B, at random, from the group. A and B play the asymmetric hawk-dove game. Each player is supposed to select hawk (escalation) or dove (retreat). If both opt for dove, they share the reward $V$ equally without fighting. Each receives $V / 2$. If one player opts for hawk and the other player opts for dove, the hawk player wins and receives reward $V$. The dove loses and gains no reward. If both opt for hawk, they fight actually. The player who wins receives the reward $V$ while the player who loses has to pay the cost, $-C$ ($V$, $C > 0$). The chance that A wins against B is determined by the difference of RHP of A and B based on the function $\theta(x_A, x_B)$ in the eq. (1) below.

$$\theta(x_A, x_B) = \frac{1}{1 + e^{-(x_A - x_B)/a}}, \quad (1)$$

where $x_A$ and $x_B$ correspond to RHP for players A and B respectively. Eq. (1) suggests that when the A's RHP is higher than B's, A is more likely to win. When the value of $a$ is lower, the probability that a player with a higher RHP would win is higher. The classical hawk–dove game assumes that $\theta(x_A, x_B)$ is $1/2$ regardless of RHP.

It is an evolutionarily stable strategy (ESS) that players opt for hawk with a chance of $V/C$ when $V < C$, or that players always opt for hawk when $V \geq C$.

## 2.2. Strategies and assumptions
### 2.2.1. Three types of inference processes

The strategies on which players select hawk or dove are genetically determined traits. We assume three types of strategies: mixer strategy (M), immediate inference strategy (II), and transitive inference strategy (TI$_{x\text{-}y}$) (Table 1). As listed in Table 1, the strategies comprise of inference and information processes. The $x$ - $y$ components in TI$_{x\text{-}y}$ represent the information processes characterized as a combination of the number of reference members and the number of shared reference members.

Each strategy consists of some of three types of inference processes, including mixer-process, immediate inference (II)-process, and transitive inference (TI)-process.

First, we will explain three types of inference processes. In mixer-process, a player makes a selection between hawk and dove following a mixed ESS where hawk is selected with a probability of $V/C$ and dove with $1 - V/C$ when $C$ (cost) $\geq V$ (reward). The player does not infer the strength of others. In addition, a player adopts the mixer-process when there is no assessment due to the lack of both ties and related contests.

We define $R_X(B|A)$ as an assessment by player X of the relative rank of player B to A based on the past interactions between A and B. $R_X(B|A)$ takes one of three values, 1,

−1 or 0. $R_X(B|A) = 1$ indicates that X assesses B stronger than A, if B has more wins than losses in the past contests between A and B. $R_X(B|A) = -1$ suggests that X assesses A stronger than B, if A has more wins than losses. $R_X(B|A) = 0$ means that X perceives A and B indifferent, if A ties with B or if there are no contests between A and B. We equally count as wins (losses) both wins (losses) in hawk vs. hawk and wins (losses) in hawk vs. dove. We consider only the signs, positive or negative, of differences of the numbers of wins and losses, not the magnitude of the differences.

In immediate inference-process, player A selects hawk when $R_A(B|A) = -1$ and dove when $R_A(B|A) = 1$. Similarly, player B opts for hawk when $R_B(A|B) = -1$ and dove when $R_B(A|B) = 1$.

With regard to the transitive inference-process, we assume that $TI_{x\text{-}y}$ players have the ability to observe and recall all contestants and results of contests only among $x$ reference members in their information set where $y$ reference members out of $x$ are shared among the $TI_{x\text{-}y}$ players.

A set of shared reference members, referred to as $y$, is randomly determined from the group. Once $y$ players are set, $(x - y)$ players are selected randomly from the group. We decide to select reference members randomly from the group, not in other ways. In fact, how to select reference members could depend on the relationships and availabilities among individuals under different social settings. Such realistic ways of selecting reference members would require more perplexing assumptions including various social contexts. Therefore, the random selection of reference members allows our study to focus on the complexity by the large group size.

Let us consider player A and B who need to assess the strengths each other. They have no direct contest, but both have direct contests with player C in the past. If $R_A(B|C) = 1$, or B > C, and $R_A(C|A) = 1$, or C > A, then $R_A(B|A)\ (= R_A(B|C) + R_A(C|A)) = 2 > 0$, or B > A. If $R_A(B|A)\ (= R_A(B|C) + R_A(C|A)) > 0$, then we set $R_A(B|A) = 1$. Here, transition inference suggests that if A < C and C < B, then A < B.

Similarly, if $R_A(B|C) = -1$, or C > B, and $R_A(C|A) = -1$, or A > C, then $R_A(B|A)\ (= R_A(B|C) + R_A(C|A)) = -2 < 0$, or A > B. If $R_A(B|A)\ (= R_A(B|C) + R_A(C|A)) <$

0, then we set $R_A(B|A) = -1$. Then, transition inference intimates that if A > C and C > B, then A > B.

If players cannot infer the strength of the opponent with transitive inference, the players follow a mixed ESS. For example, if $R_A(B|C) = 1$, or B > C and $R_A(C|A) = -1$, or A > C then A considers that B is as potent as A ($R_A(B|A) = R_A(B|C) + R_A(C|A) = 0$). If $R_A(B|A)$ ($= R_A(B|C) + R_A(C|A)) = 0$, then we set $R_A(B|A) = 0$. In this case transitive inference suggests no difference between A and B.

We introduce a function $F(R)$ defined as follows to simplify the process: $F(R) = 1$ (if $R > 0$), $F(R) = 0$ (if $R = 0$), and $F(R) = -1$ (if $R < 0$). With the function, $R_A(B|A)$ can be expressed as:

$$R_A(B|A) = F(R_A(B|C) + R_A(C|A)) \quad . \tag{2}$$

Generally, the number of opponents in common between A and B can be 2 or more. We refer to the individual common opponents as $CO_i$. We calculate $R_X(B|A)$ based on each $CO_i$. Then, transitive inference-process is defined as follows: $CO_i$ are included in a set of players in the reference members and the maximum number of $CO_i$ is $x$. The number of $CO_i$ is $n$. Therefore, $R_X(B|A)$ can be expressed as:

$$R_X(B|A) = F(\frac{1}{n} \sum_i^n F(R_X(B|CO_i) + R_X(CO_i|A))) \quad . \tag{3}$$

Using Figure 1, let us explain how player A and C, $TI_{3\text{-}2}$ players, assess RHP each other. If A and C have direct contests with A's reference members, D, E and F, player A could assess the relative rank of A to C when there are no direct contests between A and C based on eq. (3) as follows:

$R_A(C|A) = F(1/3((F(R_A(C|D) + R_A(D|A)) + F(R_A(C|E) + R_A(E|A))$
$\qquad + F(R_A(C|F) + R_A(F|A)))) \quad .$

If A does not have direct contests with F, $R_A(F|A)$ is not available. The transitive inference-process is based on the following equation, instead of the equation above:

$R_A(C|A) = F(1/2(F(R_A(C|D) + R_A(D|A)) + F(R_A(C|E) + R_A(E|A)))) \quad .$

Similarly, if C and A have direct contests with C's reference members, D, E and H, player C could assess the relative rank of C to A when there are no direct contests between the two based on eq. (3) as follows:

$R_C(A|C) = F(1/3((F(R_C(A|D) + R_C(D|C)) + F(R_C(A|E) + R_C(E|C))$

$$+ F\left(R_C(\text{A}|\text{H}) + R_C(\text{H}|\text{C})\right))) \qquad .$$

Thus, the assessment by A of relative rank of A to C through shared reference members, D and E is common with the assessment by C of relative rank of C to A. Therefore, the social hierarchies built by TI$_{x\text{-}y}$ with more shared reference members will be more similar as the number of shared reference members increases.

Our assumption allows player D to be part of $y$ players if D is also a TI$_{3\text{-}2}$ strategist, because $x$ and $y$ are assumed to be selected from a group including the focal players. In this case, we define $R_D(\text{D}|\text{D}) = 0$. In general, $R_X(\text{X}|\text{X})$ is defined as zero when X represents a player employing the TI$_{x\text{-}y}$ strategy.

In transitive inference-process, player A opts for hawk if $R_A(\text{B}|\text{A}) < 0$ and dove if $R_A(\text{B}|\text{A}) > 0$.

### *2.2.2 The definition of the strategies*

The mixer strategy always employs mixer-process and does not require information about the contests (Table 1). Immediate inference strategy uses immediate inference-process basically and then mixer-process when the immediate inference-process does not produce information useful for an assessment based on information about contests the focal players directly involved (Table 1). TI$_{x\text{-}y}$ strategy first relies on the immediate inference-process, shifts to the transitive inference-process when the immediate inference-process produces no useful information for an assessment and finally shifts to the mixer-process when the transitive inference-process results in no useful information (Table 1). Information processes for TI$_{x\text{-}y}$ strategy are based on the contests by the $x$ reference members where $y$ reference members are shared.

We focus on the situations where the group size, $N$, ranges from 10 to 50 members, large relative to the size of reference members and the cost of losing is high. This is because we consider the relative group size of cognitive abilities represented by the size of reference members, not absolute group size, is critically important in light of our research question. We defined the ranges of the number of reference members and the number of shared reference members both from 0 to 8 by 2 to facilitate the analysis of a broad range of parameters without a significant increase in computational complexity caused by an increase in the group size.

The present study employs 16 strategies in total; mixer, immediate inference, and 14 types of transitive inference strategies expressed as $TI_{x-y}$, including $TI_{2-0}$, $TI_{2-2}$, $TI_{4-0}$, $TI_{4-2}$, $TI_{4-4}$, $TI_{6-0}$, $TI_{6-2}$, $TI_{6-4}$, $TI_{6-6}$, $TI_{8-0}$, $TI_{8-2}$, $TI_{8-4}$, $TI_{8-6}$ and $TI_{8-8}$. The strategies are designed to study how transitive inference evolves under the limited social cognition as defined above.

| Strategies | Inference processes | | | Information processes | | |
|---|---|---|---|---|---|---|
| | TI-process | II-process | Mixer-process | $x$ | $y$ | MC |
| M | - | - | ✓ | - | - | - |
| II | - | ✓(1) | ✓(2) | 0 | 0 | 14 |
| $TI_{2-0}$ | ✓(2) | ✓(1) | ✓(3) | 2 | 0 | 14 |
| $TI_{2-2}$ | ✓(2) | ✓(1) | ✓(3) | 2 | 2 | 14 |
| $TI_{4-0}$ | ✓(2) | ✓(1) | ✓(3) | 4 | 0 | 14 |
| $TI_{4-2}$ | ✓(2) | ✓(1) | ✓(3) | 4 | 2 | 14 |
| $TI_{4-4}$ | ✓(2) | ✓(1) | ✓(3) | 4 | 4 | 14 |
| $TI_{6-0}$ | ✓(2) | ✓(1) | ✓(3) | 6 | 0 | 14 |
| $TI_{6-2}$ | ✓(2) | ✓(1) | ✓(3) | 6 | 2 | 14 |
| $TI_{6-4}$ | ✓(2) | ✓(1) | ✓(3) | 6 | 4 | 14 |
| $TI_{6-6}$ | ✓(2) | ✓(1) | ✓(3) | 6 | 6 | 14 |
| $TI_{8-0}$ | ✓(2) | ✓(1) | ✓(3) | 8 | 0 | 14 |
| $TI_{8-2}$ | ✓(2) | ✓(1) | ✓(3) | 8 | 2 | 14 |
| $TI_{8-4}$ | ✓(2) | ✓(1) | ✓(3) | 8 | 4 | 14 |
| $TI_{8-6}$ | ✓(2) | ✓(1) | ✓(3) | 8 | 6 | 14 |
| $TI_{8-8}$ | ✓(2) | ✓(1) | ✓(3) | 8 | 8 | 14 |

Table 1: Summary of strategies
The mark ✓ shows which inference process each strategy adopts. The number in ( ) next to ✓ indicates the order of priority in inference processes. For example, when (1) is available (1) is employed and when (1) is not available (2) is employed. 1 is the highest priority order and 3 is the lowest. MC in information processes stands for memory capacity defined as the number of contests players can remember.

In our context, standard transitive inference, which appears in Doi and Nakamaru (2018) and Nakamaru and Sasaki (2003), is considered as $TI_{N-N}$ when the group size is $N$. Standard transitive inference represents a unique case where the number of shared reference members, the number of reference members and the group size are all equal to $N$. In standard transitive inference, all players can recognize and recall all players and information about them in a group. Our study focuses on more general circumstances with the number of shared reference members ≤ the number of reference members < the group size, where players can recognize and recall only a limited number of other players in a group. $TI_{x\text{-}y}$ represents more limited information processes than $TI_{N\text{-}N}$ because $x$ and $y$ are not greater than the group size, $N$.

When the group size is smaller and closer to the number of reference members, ($x - y$) players are more likely to be overlapped among players with the same strategy $TI_{x\text{-}y}$. Before making a detailed explanation, our brief conclusion is that impacts should be very marginal when the group size is greater than 10 considering that the number of reference members is equivalent to eight. Overlapping members in a set of reference members among the same strategists in the group emerges when the number of reference members is close or equal to the group size. When the number of reference members is equal to the group size, all members in the set of reference members are identical. Therefore, all members share all reference members ($x = y$ as a result). If a set of reference members is determined randomly from the group, assuming that the number of shared reference members is zero, we can count how many members in a set of reference members may overlap. As the number of reference members decreases to a level lower than the group size, the expected number of overlapped reference members among the same strategists, declines. For example, when the group size and the number of reference members are eight, any $TI_{8\text{-}y}$ ($y < 8$) is identical to $TI_{8\text{-}8}$. When the group size is eight and the number of reference members is seven, the number of overlapped reference members declines substantially. To clarify the impacts of the overlapping, we simulated how many reference members would overlap when the group size is ten assuming that a set of reference members is each determined randomly and the number of shared reference members is zero, or $TI_{w\text{-}0}$. We observe that the number of overlapped members among all members is 10 when $w = 10$; four when $w = 9$; one when $w = 8$, and zero when $w = 7$. These results suggest that such overlapping could influence $TI_{8\text{-}y}$ ($y < 8$) marginally but would not affect any $TI_{x\text{-}y}$ ($x \leq$ the number of reference members = 7) when the group size is 10. Therefore, we do not consider that the overlapping could influence any $TI_{x\text{-}y}$ when the group size is larger than 10. Overlapping would not matter overall because we focused on a large group.

*2.3. Evolutionary dynamics with mutation*

We consider a generation of $T$ units of time. We assign a new RHP to each player at the beginning of each generation in a random manner and remains unchanged until it is reset. RHP is regarded as a nonhereditary trait expressed as a real number randomly chosen from a uniform distribution between 0 and 10, exclusive of 10. In one unit of time, two players who are randomly picked from the group play the hawk–dove game. The players opt for hawk or dove based on their strategies. After repeating the procedure $T$ times in a single generation, the payoff for players is aggregated strategy by strategy. Subsequently, the number of players with the specific strategy at the start of the next generation is proportional to the aggregate payoff of players for the strategy in the prior generation. The aggregate payoff is calculated to be positive by adding an absolute value of expected minimum payoffs to all players in order to avoid negative payoffs.

We assume that mutation occurs in the following two loci with a probability of $\mu$ independently: one is the number of reference members, referred to as $x$-locus and the other is the number of shared reference members, referred to as $y$-locus. Here, the number of reference members is $x$ and the number of shared reference members is $y$.

Even though mixer and immediate inference strategies do not depend on the number of reference members or the number of shared reference members, we technically assign $x = 0$ to the mixer strategy, $x = 1$ to the immediate inference strategy, and $y = 0$ to both mixer and immediate inference strategies. Then combinations of $x$ and $y$ are unique to each strategy so that mutation in the $x$ and/or $y$ loci means mutation in strategies.

We assume that mutation is allowed to occur randomly in the $x$-locus and then in the $y$-locus regardless of the current positions in the arrays. The new values in the $x$-locus and in the $y$-locus following mutation are allowed to adopt any values in the $x$-locus and the $y$-locus under $y \leq x$ conditions. So, $x \in \{0, 2, 4, 6, 8\}$. For each $x$, $y \in \{0\}$ in $x = 0$, $y \in \{0, 2\}$ in $x = 2$, $y \in \{0, 2, 4\}$ in $x = 4$, $y \in \{0, 2, 4, 6\}$ in $x = 6$, and $y \in \{0, 2, 4, 6, 8\}$ in $x = 8$. For example, when the prevailing positions in $x$-locus and $y$-locus are 2 and 0, respectively, the new $x$-locus value following mutation could be 0, 4, 6 or 8, excluding 2, the current value, with the same probability, $\mu / 4$. If the new value in the $x$-locus is 8, the new $y$-locus values could be 2, 4, 6 or 8 excluding 0, the prevailing value, with the same probability, $\mu / 4$.

Finally, the next generation begins. This process continues over $G$ generations. The group size is constant throughout a generation. Here we apply $\mu = 0.001$.

*2.4.  Key parameters*

There are four key parameters characterizing social conditions including 1) group size ($N$), 2) $C / V$ ratio, which is a cost divided by a reward, 3) $N_p$ (= $2T / (N \times (N - 1))$), referring to the expected number of contests per pair of players, and 4) Memory capacity ($MC$).

Here we use $N_p = 2$ because $N_p = 2$ gives two chances of participating in a contest to any pairs on average and Doi and Nakamaru (2018) suggest that TI works well under $N_p = 2$. $N_p = 2$ means that the encounter rates remain constant regardless of the group size because we increase $T$ units of time as the group size increases. We use the constant $N_p = 2$ for all analyses in the present study for simplicity.

In the present study, we consider the group size ($N$) as one of components of social complexity as discussed in Section 1.2.

How reliable information from contests is in assessing RHP depends on the $C / V$ ratio. For example, the probability ($= (V / C)^2$) that both players opt for hawk is low when $C / V$ is high, so that results do not reflect actual RHP because the rank is often set without actual fights. The $C / V$ ratio is a key parameter influencing what strategies can persist. We maintain the reward constant ($V = 4$) and vary the cost. We focus on the results when the cost is high ($C = 30$) because it is known that transitive inference persists in high-cost environments (Nakamaru and Sasaki, 2003).

Memory capacity ($MC$) is defined as the number of contests players can remember. For example, immediate inference players maintain $MC$ of records in memory about contestants and the results of their own direct contests. We assume that players forget older records beyond memory capacity and maintain only the latest $MC$ of records. In the present study, we apply a constant memory capacity ($MC = 14$) for all analyses because we consider it reasonable to assume that memory capacity is limited. The minimum memory capacity required for an individual to understand a relationship with others is $N - 1$. We consider $N - 1$ too low as a memory capacity; therefore, we set memory capacity as $2 \times (N - 1)$ given $N_p = 2$. $MC = 14$ assumes that the lowest size of a group is eight. When the group size is eight, TI$_{8-8}$ with $MC = 14$ represents adequate social cognition. This assumption means that individuals can remember 14 records of contests out of the expected numbers of encounters, 98 (= 2 × (50 − 1)), when *N* is 50. All observations in memory are treated equally. However, inference process in the strategies gives priority to information about direct contests by first applying immediate inference-process, which is more direct experiences and then transitive inference-process in case of no direct contests.

## 3. Results

We explored the evolutionary dynamics of strategies in various group sizes. We ran the evolutionary simulations with mutation with all 16 strategies over 10,000 generations, iterated 50 times, and calculated the average of population frequencies at each generation for each strategy. As all strategies converges to a single strategy over the generations in a single run except a small number of mutants, each run ends up with the most dominant strategy without coexistence of strategies (Fig. 2 (a)). We assumed that an initial strategy for all players is a mixer strategy. Average final frequencies of the strategies after 50 runs are presented in Figure 2 (b). Average final frequencies in Figure 2 (b) mean how often each most dominant strategy appears in the 50 runs.

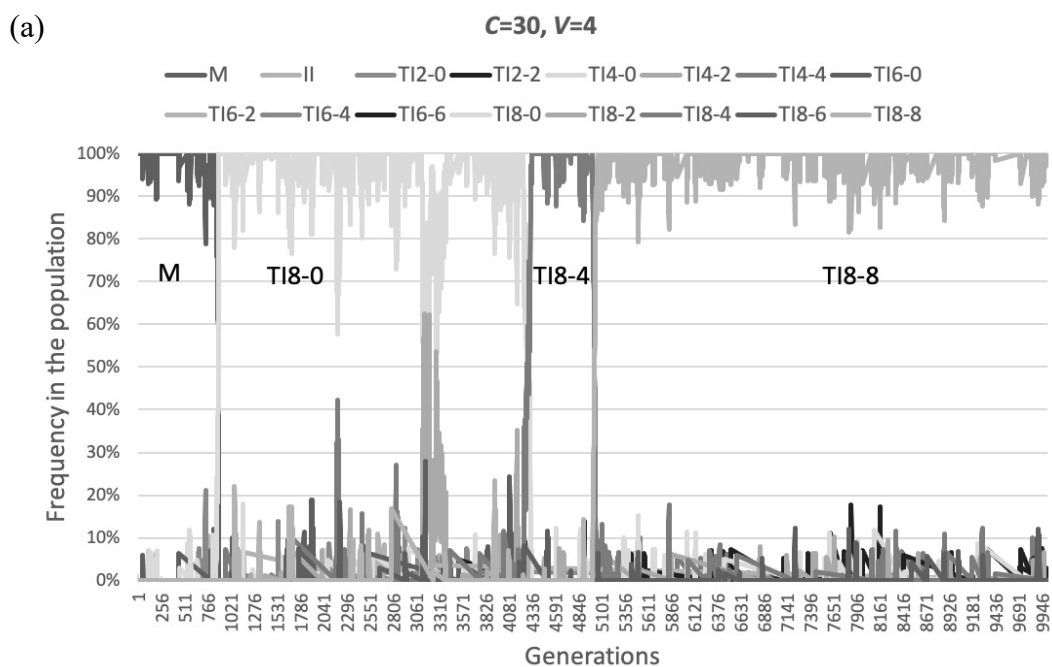

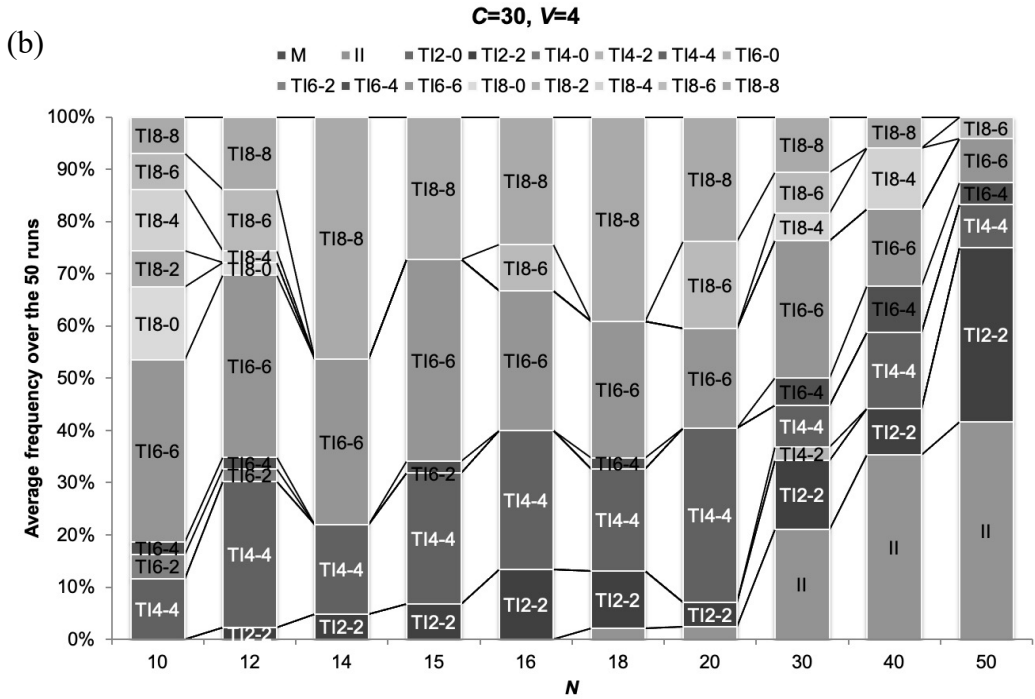

Fig. 2: Evolutionary simulation with random mutation
We examined evolutionary dynamics of all strategies with mutation that occurs in two loci with a probability of $\mu$ (= 0.001) independently: one is for the number of reference members and the other is for the number of shared reference members. Initial strategy for all players is always a mixer strategy. Here we use the number of generations ($G$) = 10 000, $\mu$ = 0.001, $MC$ = 14, $N_p$ = 2, $V$ = 4 and $C$ = 30.
(a) The vertical axis represents the final frequencies of strategies over generations ($G$) in a single run and the horizontal axis represents $G$. Here we use $N$ = 20.
(b) The vertical axis represents the final frequencies of strategies as averages over 50 iterations and the horizontal axis represents $N$.

Our analysis confirms that transitive inference strategies are collectively more dominant than the immediate inference strategy across any group sizes (Fig. 2 (b)). Concretely TI$_{Z\text{-}Z}$ ($Z$ = 2, 4, 6 and 8) strategies turn out to be more successful than other TI strategies, TI$_{Z\text{-}Y}$ ($Y < Z$). We also note that immediate inference strategy becomes more successful when $N \geq 30$.

What if we introduce the standard transitive inference strategy, TI$_{N\text{-}N}$ where $N$ is the group size? We looked into the evolutionary dynamics of strategies of M, II, TI$_{2\text{-}2}$, TI$_{4\text{-}4}$, TI$_{6\text{-}6}$, TI$_{8\text{-}8}$ and TI$_{35\text{-}35}$ under $N$ = 35 and $MC$ = 14. We assume that all strategies start with equal initial frequencies and no mutation occurs over 500 generations and repeated it 50 times. As all strategies converges to a single strategy over the generations in a single run except a small number of mutants (Fig. 2 (a)), each run ends up with the most dominant strategy without coexistence of strategies. Average final frequencies of the strategies after the 50 runs are presented in Figure 3. Average final frequencies in Figure 3 mean how often each most dominant strategy appears in the 50 runs.

Results in Figure 3 show that final frequencies of $TI_{4-4}$, $TI_{6-6}$, $TI_{8-8}$ and $TI_{35-35}$ are similar, and $TI_{2-2}$ ends up with the smaller final frequency. Immediate inference strategy does not survive in Fig. 3 while it succeeds in Fig. 2 (b) when $N \geq 30$. As we will explain more closely later in this section, it is because $TI_{Z-Z}$ is less likely to survive when the initial frequency of $TI_{Z-Z}$ is smaller. All strategies including $TI_{N-N}$ start with equal frequencies at the start in Fig. 3 while initial frequencies of $TI_{Z-Z}$ are set zero at the start in Fig. 2 (b). It is noteworthy that $TI_{35-35}$, the standard transitive inference that can use information of all members in the group, does not make a meaningful difference from $TI_{Z-Z}$ ($Z = 4$, 6 and 8), under limited memory capacity that prevents the standard transitive inference from storing information about all other members. It is true that $TI_{Z-Z}$ ($Z = 4$, 6, 8 and $N$) is more successful than $TI_{2-2}$ consistently both in Fig. 2 (b) and 3. Results in Fig. 2 (b) and 3 suggest that the size of $Z$ is not a key factor for survival as long as $Z$ is greater than 2.

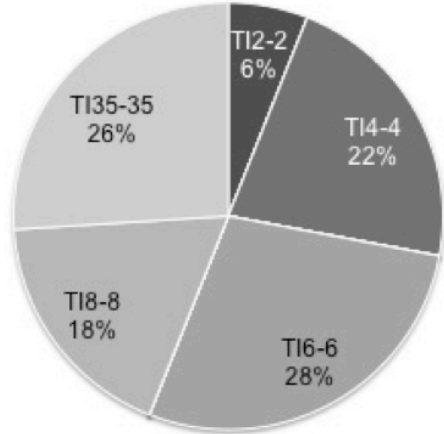

Fig. 3: Evolutionary simulation without random mutation
We analyzed evolutionary dynamics of M, II, $TI_{2-2}$, $TI_{4-4}$, $TI_{6-6}$, $TI_{8-8}$ and $TI_{35-35}$ with equal initial proportions under $N = 35$. No mutation is assumed. We ran the process 50 times and calculated the average frequency of each strategy. Each run ended with 100% of the most dominant strategy and there was no coexistence of strategies. Final strategy frequencies represent how often the respective strategies become the most dominant strategy. We calculated the averages of the final frequencies only when the survival strategy converged into a single strategy. Here $G = 500$, $MC = 14$, $N_p = 2$, $V = 4$ and $C = 30$.

What is a key factor if the size of $Z$ is not a key factor? We examined how consistently social hierarchies are built in $TI_{Z-Z}$ and $TI_{Z-Y}$ ($Y < Z$) under $Z = 8$. Figure A1 demonstrates that $TI_{Z-Z}$ strategies can form the social hierarchy better than $TI_{Z-Y}$ ($Y < Z$) under any $C$. More importantly, this result suggests that sharing reference members more with other members promotes the prompt establishment of the social hierarchy by using information from others' experiences. The key factor to succeed is not the number of reference members, but the ability to share the reference members (Fig. A1). In addition, figure A2 shows that TI$_{Z-Z}$ ($Z$ = 2, 4, 6, 8 and $N$) strategies can form the social hierarchy faster than immediate inference strategy. This promotes the evolution of transitive inference more in larger $C$. Forming the linear social hierarchy is more important for survival in large $C$.

Why does most TI$_{Z-Z}$ strategies succeed similarly despite different $Z$? We examined how quickly the social hierarchies are built in II and TI$_{Z-Z}$ ($Z$ = 2, 4, 6, 8 and $N$) and found that $CI_1$ in all TI$_{Z-Z}$ ($Z$ > 2) also develops indifferently (Fig. A2). We consider that similar $CI_1$ behaviors are the reason for similar success in TI$_{Z-Z}$ ($Z$ > 2).

However, this finding seems a little counter-intuitive because higher $Z$ should suggest higher cognitive abilities. We look into how $CI_1$ in all TI$_{Z-Z}$ ($Z$ > 2) develops under unlimited memory capacity and confirm that $CI_1$ with higher $Z$ increases faster (Fig. A3). This means that limited memory capacity prevents TI$_{Z-Z}$ with higher $Z$ from being more successful. We find that all TI$_{Z-Z}$ ($Z$ > 2) behaves similarly because of limited memory capacity. Importantly $Z$ does not make a difference under limited memory capacity.

TI$_{Z-Z}$ strategies dominate TI$_{Z-Y}$ ($Y$ < $Z$) strategies and TI$_{Z-Z}$ even with the smallest $Z$ survives broadly across various group sizes (Fig. 2 (b)). This suggests that even limited social cognition that includes the ability in transitive inference to observe interactions among others works better than social cognition in immediate inference that does not have the ability to observe interactions among others. TI$_{Z-Z}$ strategies dominate TI$_{Z-Y}$ ($Y$ < $Z$) strategies because sharing reference members more with other members promotes the prompt formation of the social hierarchy by using information from others' experiences (Fig. 2 (a) and Fig. A1). The ability to share reference members is more important than the ability to broaden a set of reference members especially when memory capacity is limited (Fig. 2 (b) and A3). The evolutionary simulations beginning with all players applying TI$_{8-8}$, TI$_{4-4}$, or immediate inference end up with all players maintaining their respective strategies even at the end in a large group ($N$ = 40) (Table 2). TI$_{8-8}$, TI$_{4-4}$ and immediate inference are all evolutionarily stable and could evolve if they are applied by the majority of a group. On the other hand, the evolutionary simulations beginning with all players applying TI$_{8-0}$ or TI$_{4-0}$ end with various combinations of final frequencies of different strategies (Table 2-D and E). It is confirmed that TI$_{8-0}$ and TI$_{4-0}$ are not ESSs. In sum, TI$_{Z-Z}$ ($Z$ < group size ($N$)) is an ESS while TI$_{Z-0}$ is not an ESS because TI$_{Z-Z}$ shares reference members with others while TI$_{Z-0}$ does not. As discussed earlier, the ability to share reference members is critical because it facilitates the prompt establishment of the social hierarchy (Fig. A1 and A2).

| | Frequencies | M | II | TI$_{2-0}$ | TI$_{2-2}$ | TI$_{4-0}$ | TI$_{4-2}$ | TI$_{4-4}$ | TI$_{6-0}$ | TI$_{6-2}$ | TI$_{6-4}$ | TI$_{6-6}$ | TI$_{8-0}$ | TI$_{8-2}$ | TI$_{8-4}$ | TI$_{8-6}$ | TI$_{8-8}$ |
|---|---|---|---|---|---|---|---|---|---|---|---|---|---|---|---|---|---|
| A | Initial | | | | | | | | | | | | | | | | 1.0 |
| A | Final | | | | | | | | | | | | | | | | 1.0 |
| B | Initial | | | | | | | 1.0 | | | | | | | | | |
| B | Final | | | | | | | 1.0 | | | | | | | | | |
| C | Initial | | 1.0 | | | | | | | | | | | | | | |
| C | Final | | 1.0 | | | | | | | | | | | | | | |
| D | Initial | | | | | 1.0 | | | | | | | | | | | |
| D | Final | | | | | | | 0.7 | | | | 0.1 | | | | | 0.1 |
| E | Initial | | | | | | | | | | | | 1.0 | | | | |
| E | Final | | | | | | | 0.1 | | | | 0.1 | | | | 0.2 | 0.6 |

Table 2: Evolutionary dynamics of all strategies with the random mutations that take place in two loci with a probability of $\mu$ (= 0.001) independently; one is for $x$ and the other is for $y$ in TI$_{x-y}$. Each case, A, B, C, D, and E has a different initial strategy frequency. Initial strategy frequencies are as follows; A with TI$_{8-8}$ = 100%, B with TI$_{4-4}$ = 100%, C with II = 100%, D with TI$_{4-0}$ = 100% and E with TI$_{8-0}$ = 100%. Numbers in each cell represent the strategy frequencies at the start (upper row) and the end (lower raw) for each case, as averages over 50 times. Each run ends up with 100% of the most dominant strategies and no coexistence of strategies. Final strategy frequencies represent how often the respective strategies become the most dominant strategy. We calculate an average of final frequencies only when the survival strategy converges into a single strategy. The numbers in cells are rounded and sum of the numbers may not be 1 because of the rounding. We examine cases with two different $C/V$ ratios (1.25 and 4). Here we use $N$ = 40, $G$ = 10 000, $\mu$ = 0.001, $MC$ = 14, $C$ = 30 and $V$ = 4.

As mentioned earlier, why does immediate strategy start to appear again and TI$_{Z-Z}$ with higher $Z$ begins to dominate less when the group size becomes very large ($N \geq 30$) (Fig. 2 (b))? We consider that one of reasons is that the success of TI$_{Z-Z}$ depends on initial proportions of strategies. TI$_{Z-Z}$ with higher $Z$ may require a higher initial proportion. We examined the evolutionary dynamics existing between immediate inference and TI$_{Z-Z}$ under different group sizes to observe how final frequencies of TI$_{Z-Z}$ develop over immediate inference with an increase in the group size (Fig. 4). No mutation was assumed. Figure 4 shows that TI$_{Z-Z}$ strategies with higher (lower) initial proportions tend to result in higher (lower) final frequencies. This result suggests that TI$_{Z-Z}$ has dependency on the initial proportions, meaning that TI$_{Z-Z}$ requires a larger number of players following the same strategy to recognize the similar hierarchy. TI$_{Z-Z}$ strategies survive over immediate inference in Fig. 3 where all strategies including TI$_{N-N}$ start with equal frequencies at the start, because TI$_{Z-Z}$ is more likely to survive when the initial frequency of TI$_{Z-Z}$ is greater.

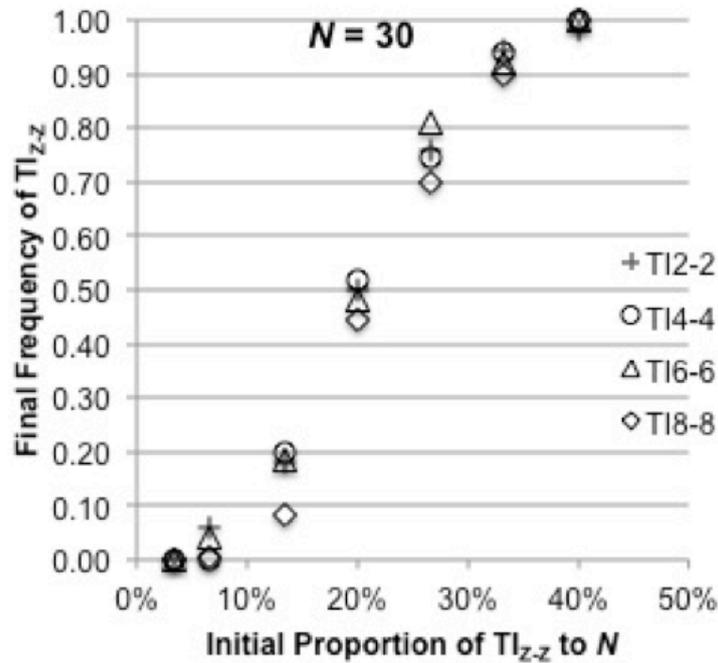

Fig. 4: Influence of initial population on TI$_{Z-Z}$
We analyzed evolutionary dynamics between II vs. TI$_{Z-Z}$ with various initial proportions of TI$_{Z-Z}$ under $N$ = 30. $Z$ = 2, 4, 6 and 8. The vertical axis represents the final frequencies of TI$_{Z-Z}$ as averages over 50 iterations and the horizontal axis represents initial proportions of TI$_{Z-Z}$ as % share of an entire population. Here $MC$ = 14, $N_p$ = 2, $V$ = 4 and $C$ = 30.

We consider that the other reason is that, as this study assumes limited cognitive abilities, differences in all strategies become meaningfully small as $N$ becomes very large ($N \geq 30$). For example, the TI$_{8-8}$ strategy with the largest cognitive abilities can observe only 8 members in the group with more than 30 members. We looked into how $CI_1$ between TI$_{2-2}$ and TI$_{Z-Z}$ ($Z > 2$) develop under different $N$ and find that $CI_1$ between TI$_{2-2}$ and TI$_{Z-Z}$ ($Z > 2$) becomes closer when $N$ exceeds 30 (Fig. A4).

We conclude that these are reasons why immediate inference strategy starts to appear and TI$_{Z-Z}$ with higher $Z$ begins to dominate less when $N$ exceeds 30.

# 4.     Discussion and conclusions

The ability to establish the social hierarchy is critical in complex societies (Hotta et al., 2015; Mikolasch et al., 2013). Furthermore, the establishment of the social hierarchy requires advanced social cognition that facilitates the identification of other members broadly, recognition and the recalling of relationships with and among other members (Bshary and Brown, 2014; Seyfarth and Cheney, 2015).

What types of social cognition and what level of social cognition are required under transitive inference? As the group size increases, the relationships among members increasingly become complex and information required for understanding the social hierarchy significantly increase. Transitive inference based on all potential group members requires more developed cognitive abilities especially as the group size increases. Instead, as heuristic approaches, animals may apply reference transitive inference that allows members to recognize and remember social interactions only among a set of reference member rather than all potential members. Animals may apply heuristic approaches such as the reference transitive inference to handle such complex scenarios, instead of developing social cognition accordingly. The reference transitive inference based on a set of reference members may produce smaller amount of information than transitive inference based on all group members. However, the reference transitive inference can build social hierarchy as quickly as standard transitive inference with much less developed cognitive abilities than standard transitive inference.

Therefore, in the present study we consider situations where individuals in a group apply information based only on relationships with reference members who are randomly selected from a group as opposed to that based on all potential members. Our study assumed that social cognition is a set of processes to a) make inferences (Inference processes) and b) to gather and store information for inference (information processes). Inference processes consist of three components, including 1) the number of reference members on, 2) the number of shared reference members, and 3) memory capacity. Two former parts correspond to heuristic mechanisms in transitive inference.

We examined how information processes in transitive inference operate in a large group.

Our study demonstrates that TI$_{Z-Z}$ could survive over immediate inference in the analysis of evolutionary dynamics with mutation in large group sizes under relatively high costs, regardless of the value of $Z$ except that $Z = 2$ (Fig. 2 (b)). TI$_{Z-Z}$ are various based on different $Z$s and one of TI$_{Z-Z}$ is the most dominant strategy in a group. Which TI$_{Z-Z}$ dominates the most in a group depends on its initial population. More importantly, the ability to share reference members is critical (Fig. 2 (a), (b) and Fig. A1) because it facilitates the prompt establishment of the social hierarchy especially when memory capacity is limited (Fig. A1, A2 and A3). On the other hand, the larger numbers of reference members lead to higher $CI_1$ under unlimited memory capacity (Fig. A3) than under $MC = 14$ (Fig. A1, A2 and A4). Doi and Nakamaru (2018) also suggests that smaller memory leads to lower $CI_1$ when $Z = N$. It is important that higher $Z$ does not pay off under limited memory capacity.

Transitive inference has been reported to evolve in animals living in large group as a way of understanding social hierarchy without increasing memory capacity with a corresponding increase in the number of dyadic relationships as the group size increases (Mikolasch et al., 2013; Paz-Y-Miño et al., 2004; Tibbetts et al., 2019). Doi and Nakamaru (2018) theoretically demonstrated that, even with limited memory capacity, transitive inference persists over immediate inference when the cost of losing is relatively high.

The findings of the present study demonstrate that transitive inference can evolve with heuristics, when the cost of losing is relatively high. This observation is potentially inconsistent with the idea that more highly developed social cognition needs to evolve as the group size increases because the larger group size increases social complexity substantially. However, the ability to share as well as have reference members with others makes a significant difference between immediate inference and transitive inference even though the number of reference members is low as far as the ability to share reference members is high. It is because, in immediate inference, information is limited to individual experiences while transitive inference with a set of reference members can apply information gathered from relationships and interactions with reference members. Overall, the results suggest that animals may apply a type of shortcut, or heuristics, in order to deal with increasing social complexity with an increase in the group size, instead of developing very high levels of social cognition.

As observed in the present study, transitive inference triumphs over immediate inference at the cost of establishing the social hierarchy rapidly rather than consistently with RHP ranks. In future studies, we will investigate what kinds of social cognition could improve the negative relationship between the speed of establishing a hierarchy and the consistency of the hierarchy with RHP.

In this research we study the evolution of transitive inference with a group of reference members as heuristics. As discussed earlier, there seems to be a positive relationship between the number of reference members and memory capacity. Therefore, in the future, we like to study the coevolution between the ability to increase the number of reference members and memory capacity.

Finally, our study theoretically proves that the reference transitive inference based on reference members works well as heuristics instead of developing cognitive abilities highly in animals in living in a large group. We will study how a set of reference members can develop in a large group.

**Appendix:** Consistency index (*CI*) provides useful information on how rapidly each strategy can form a social hierarchy.

We introduce an analytical index modified based on consistency index (*CI*) applied in Doi and Nakamaru (2018). In Doi and Nakamaru (2018), *CI* is defined as an indicator of how consistency between $R_i(j|i)$ and $R_j(i|j)$ in any two players, *i* and *j*, evolves as players play games more, assuming all players follow the same strategy in a group. Details about *CI* are discussed in Doi and Nakamaru (2018). In short, *CI* = 0 indicates that complete consensus is built where all combinations of tactics are hawk vs. dove or dove vs. hawk. Higher *CI* suggests more disagreements. The highest *CI* is 0.5, indicating complete disagreements.

In the present study we define $CI_1$ as $1 - CI / 0.5$, where $CI_1 = 1$ indicates perfect consensus while $CI_1 = 0$ means no consensus. As players play games more and more, $CI_1$ ($0 \leq CI_1 \leq 1$) is expected to increase as a social hierarchy is established.

Using $CI_1$, we investigate how the number of reference members and the number of shared reference members influences the process of establishment of social hierarchy under *MC* = 14. We examine how $CI_1$, an indicator of how rapidly each strategy can facilitate the establishment of a social hierarchy using games within a single generation for each strategy. We conduct the analysis for immediate inference, $TI_{2\text{-}2}$, $TI_{4\text{-}4}$, $TI_{6\text{-}6}$, $TI_{8\text{-}8}$ and $TI_{N\text{-}N}$ (*N* = group size = 16) strategies with the number of reference members equivalent to the number of shared reference members under three different social conditions when the cost is 30. We assume that every player follows the same strategies.

First, important result is that $TI_{Z\text{-}Z}$ is better than $TI_{Z\text{-}Y}$ ($Y < Z$) in terms of the level of $CI_1$, suggesting that $TI_{Z\text{-}Z}$ is more powerful in building the social hierarchy than $TI_{Z\text{-}Y}$ (Fig. A1).

Second Fig. A2 demonstrates that $CI_1$ in all TI strategies increases more rapidly than $CI_1$ in immediate inference strategy and $CI_1$ in $TI_{4\text{-}4}$, $TI_{6\text{-}6}$ $TI_{8\text{-}8}$ and $TI_{N\text{-}N}$ increases faster than $CI_1$ in $TI_{2\text{-}2}$ regardless of costs. Collectively $CI_1$ in $TI_{Z\text{-}Z}$ performs better than $CI_1$ in immediate inference, which suggests that TI with the smallest *Z* contributes to the more rapid establishment of the social hierarchy than immediate inference. The finding that $CI_1$ in $TI_{Z\text{-}Z}$ ($Z > 2$) is better than $CI_1$ in $TI_{2\text{-}2}$ suggests that the number of reference members and shared reference members needs to be large to some extent. On the other hand, it seems a little counter-intuitive that $CI_1$ in TI $_{Z-Z}$ ($Z > 2$) behaves very similarly despite expected difference in their cognitive abilities because of different $Z$. This is a kind of puzzle.

To solve the puzzle, we looked into how $CI_1$ in in TI $_{Z-Z}$ develops under unlimited memory capacity instead of $MC = 14$. We confirm that $CI_1$ in TI$_{Z-Z}$ with higher $Z$ increases higher under unlimited memory capacity (Fig. A3). Expectedly higher cognitive abilities with higher $Z$ improve the abilities to form the social hierarchy. This suggests that, under limited memory capacity, having broader reference members does not necessarily lead to the prompter formation of the social hierarchy. In contrast, sharing more reference members is more important than broadening a set of reference members (Fig. A1 and A3).

Finally given $Z \leq 8$ assumed in the present study, as the group size ($N$) increases differences of $CI_1$ developments with different $Z$ (= 2, 4, 6 and 8) is supposed to be more marginal (Fig. A4). This is especially true between TI$_{2-2}$ and TI$_{Z-Z}$ ($Z > 2$) (Fig. A4).

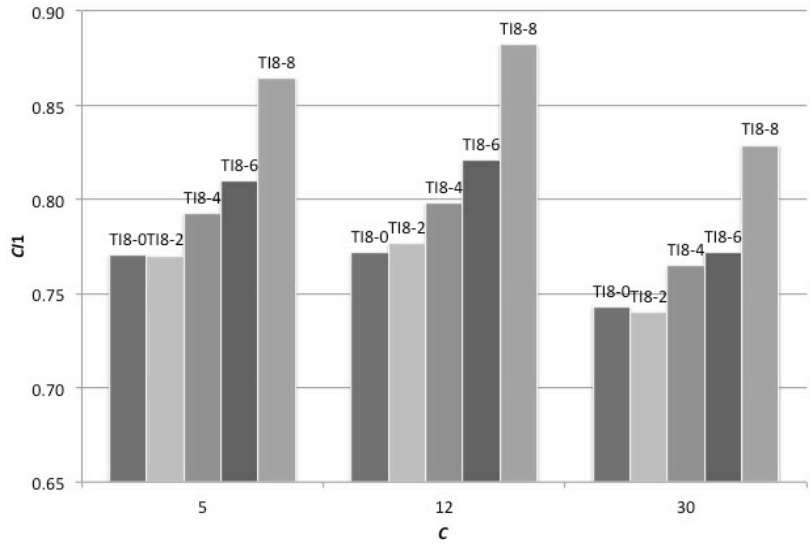

Fig. A1: $CI$ level based on strategies with a constant number of reference members and different numbers of shared reference members
We ran 240 games in one generation with $N = 16$ ($N_p = 2$) and $MC = 14$. The horizontal axis indicates $C$. The vertical axis represents average $CI_1$ indices over 100 iterations. Here $V = 4$, $C = 5$, 12 and 30.

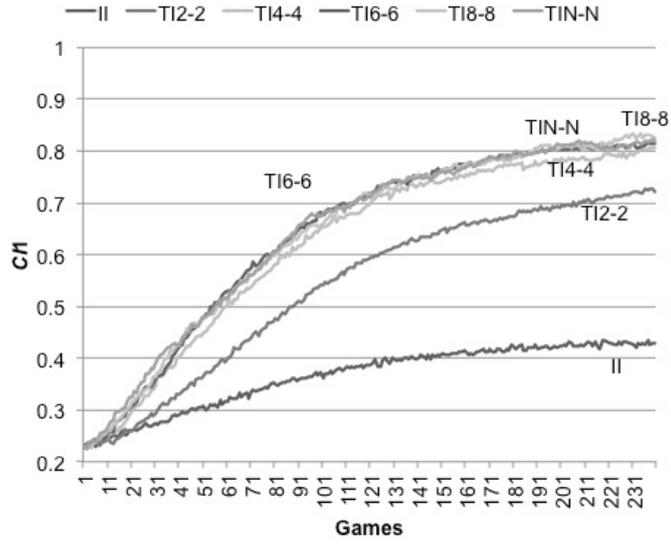

Fig. A2: $CI_1$ developments by strategies
We ran 240 games in one generation with $N = 16$ ($N_p = 2$) and $MC = 14$. The horizontal axis indicates the number of games. The vertical axis represents averages of $CI_1$ indices over 100 iterations. Line legend shows a strategy name. $C = 30$. Here $V = 4$.

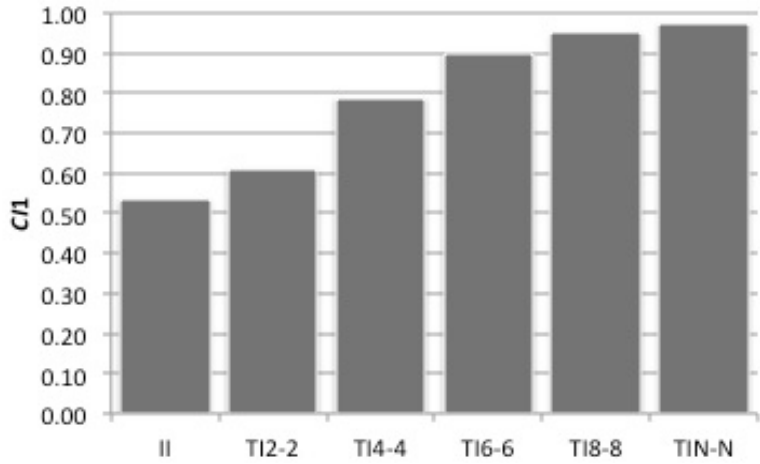

Fig. A3: $CI_1$ developments in TI$_{Z\text{-}Z}$ with unlimited memory capacity
We ran 210 games in one generation with $N = 15$ ($N_p = 2$) under unlimited memory capacity. The horizontal axis indicates a strategy name. The vertical axis represents averages of $CI_1$ indices over 50 iterations. $Z = 2, 4, 6$ and $8$. $C = 30$. Here $V = 4$.

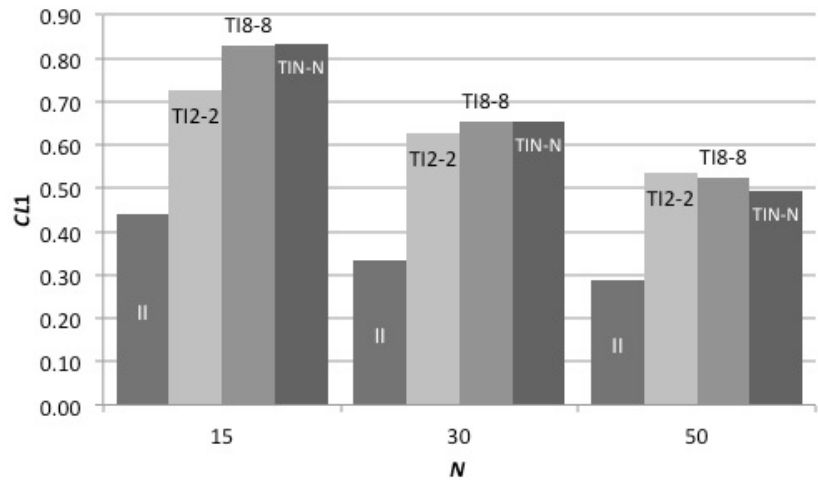

Fig. A4: $CI_1$ developments in $TI_{Z\text{-}Z}$ under different group sizes
We ran 210 games in one generation with $N = 15$ ($N_p = 2$) and with $MC = 14$. The horizontal axis indicates group size. The vertical axis represents averages of $CI_1$ indices over 50 iterations for $N = 15$ and 30 and 30 iterations for $N = 50$. $Z = 2, 4, 6$ and $8$. $C = 30$. Here $V = 4$.

**Acknowledgement**

Authors would like to thank Enago (www.enago.jp) for the English language review.

**Funding**

This work received supports by JSPS KAKENHI [grant number JP26440236 (MN)].

**Declaration of interest**

None

**Authors' contributions**

KD conceived and designed the study, analyzed the model, carried out the numerical simulations, and drafted the manuscript; MN helped to develop the study and drafted the manuscript; All the authors gave final approval for publication.